\documentclass[aps,prl,twocolumn,superscriptaddress,showpacs]{revtex4}

\usepackage{graphicx}
\usepackage{dcolumn}

\begin{document}


\title{4$f$-derived Fermi Surfaces of CeRu$_2$(Si$_{1-x}$Ge$_{x}$)$_2$ near the Quantum Critical Point: Resonant Soft X-ray ARPES Study}



\author{T.~Okane}
  \email{okanet@spring8.or.jp}
\author{T.~Ohkochi} 
\author{Y.~Takeda} 
\author{S.-i.~Fujimori}
\author{A.~Yasui} 
\author{Y.~Saitoh} 
\address{
Synchrotron Radiation Research Center, Japan Atomic Energy Agency,  Hyogo 679-5148, Japan
}%

\author{H.~Yamagami}
\affiliation{
Synchrotron Radiation Research Center, Japan Atomic Energy Agency, Hyogo 679-5148, Japan
}%
\affiliation{
Department of Physics, Kyoto Sangyo University, Kyoto 603-8555, Japan
}%

\author{A.~Fujimori}
\affiliation{
Synchrotron Radiation Research Center, Japan Atomic Energy Agency, Hyogo 679-5148, Japan
}%
\affiliation{
Department of Physics, University of Tokyo, Tokyo 113-0033, Japan
}%

\author{Y.~Matsumoto}
\author{M.~Sugi}
\author{N.~Kimura}
\author{T.~Komatsubara}
\author{H.~Aoki}
\affiliation{
Graduate School of Science and Center for Low Temperature Science, Tohoku University, Miyagi 980-8577, Japan
}%


\date{\today}

\begin{abstract}

Angle-resolved photoelectron spectroscopy in the Ce $3d$$\rightarrow$4$f$ excitation region was measured for the paramagnetic state of CeRu$_2$Si$_2$, CeRu$_2$(Si$_{0.82}$Ge$_{0.18}$)$_2$, and LaRu$_2$Si$_2$ to investigate the changes of the 4$f$ electron Fermi surfaces around the quantum critical point.    
While the difference of the Fermi surfaces between CeRu$_2$Si$_2$ and LaRu$_2$Si$_2$ was experimentally confirmed, 
a strong 4$f$-electron character was observed in the band structures and the Fermi surfaces of CeRu$_2$Si$_2$ and CeRu$_2$(Si$_{0.82}$Ge$_{0.18}$)$_2$, consequently indicating a delocalized nature of the 4$f$ electrons in both compounds.    
The absence of Fermi surface reconstruction across the critical composition suggests that SDW quantum criticality is more appropriate than local quantum criticality in CeRu$_2$(Si$_{1-x}$Ge$_{x}$)$_2$.  

\end{abstract}


\maketitle


Heavy-Fermion (HF) metals have attracted much attention in recent years as a prototypical system to study quantum criticality \cite{GEGENWART}.  
In rare-earth-based HF metals, e.g., Ce compounds, the ground state is considered to change from nonmagnetic to magnetic state at a boundary called quantum critical point (QCP) as a function of the relative strengths of Ruderman-Kittel-Kasuya-Yosida (RKKY) interaction to Kondo effect, 
and at the same time the Ce 4$f$ electrons are transformed from itinerant to localized ones. 
CeRu$_2$(Si$_{1-x}$Ge$_{x}$)$_2$ is an ideal system to study this QCP phenomena.  
CeRu$_2$Si$_2$ is a representative HF compound with a paramagnetic  ground state,  
exhibiting a metamagnetic transition at a magnetic field of $H_\mathrm{m}$ = 7.7 T.  
The large specific heat coefficient $\gamma \sim$ 350 mJ/mol K$^2$ \cite{BESNUS} indicates the itinerancy of the Ce 4$f$ electrons in the  ground state,    
and the Kondo crossover temperature $T_\mathrm{0}$ is estimated to be $\sim$ 20-25 K \cite{LOIDL,FISHER}.    
A magnetic  ground state appears upon substitution of Ge atoms for Si atoms, corresponding to the application of a negative chemical pressure:
The system is antiferromagnetic for $x$ = 0.07-0.57, ferromagnetic for $x$ = 0.57-1.0 \cite{HAEN,SUGI}, and thus the critical composition $x_{c} = 0.07$.  

If the 4$f$ electrons of the nonmagnetic Ce compound like CeRu$_2$Si$_2$ are itinerant and participate in the formation of Fermi surfaces (FSs), the total volume of the FSs should differ from that of the corresponding La compound, LaRu$_2$Si$_2$, owing to the contribution of one 4$f$ electron per unit cell.   
This FS variation is accompanied by the variation of the band structures near the Fermi level ($E_\mathrm{F}$) through hybridization between the Ce 4$f$ electrons and conduction electrons, i.e., the formation of $c$-$f$ hybridized bands, and in this sense, one can say that the energy bands have a 4$f$ electron character. 
According to band structure calculations on CeRu$_2$Si$_2$ and LaRu$_2$Si$_2$ \cite{YAMAGAMI2}, when the Ce 4$f$ electrons of CeRu$_2$Si$_2$ are assumed to be itinerant, the formation of the $c$-$f$ hybridized bands leads to the emergence of FS which includes a contribution  from heavy 4$f$ electrons.  
The de Haas van Alphen (dHvA) experiment for CeRu$_2$Si$_2$ \cite{AOKI} indicated the existence of the heavy-electron FS branches, which would correspond to the above-mentioned 4$f$-originated FS.   
On the other hand,  the dHvA experiments indicated that the FSs of ferromagnetic CeRu$_2$Ge$_2$ are almost the same as those of LaRu$_2$Ge$_2$, indicating that the 4$f$ electrons are localized in CeRu$_2$Ge$_2$ \cite{KING,IKEZAWA,YAMAGAMI1}.   
Therefore, the Ce 4$f$ electrons turn from itinerant to localized ones in going from CeRu$_2$Si$_2$ to  CeRu$_2$Ge$_2$, and the experimental determination whether the FSs of Ce-based compounds are the same or not as those of corresponding La compound is directly connected with the fact whether the Ce 4$f$ electrons are itinerant or localized.  

\begin{figure}
\includegraphics[width=6.5cm]{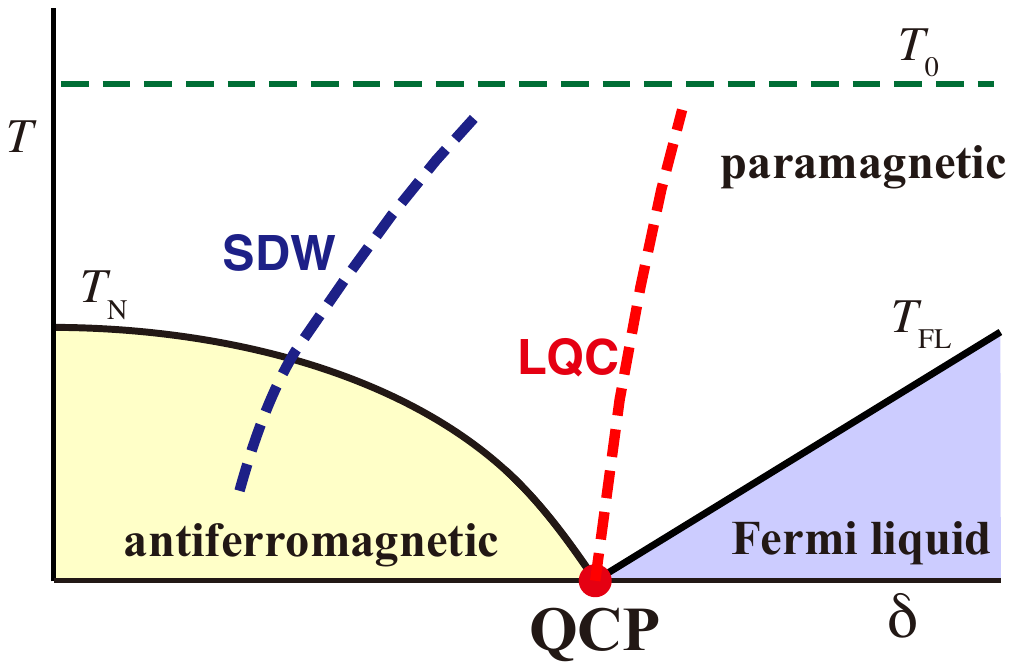}%
\caption{\label{PD}
(color online)
Schematic phase diagram for QCP of HF metal \cite{GEGENWART}.  
It is illustrated against temperature versus control parameter $\delta$ which represents the strength ratio of Kondo effect to RKKY interaction. 
$T_\mathrm{N}$, $T_\mathrm{FL}$, and $T_\mathrm{0}$ designate the N\'{e}el temperature, the onset temperature of Fermi liquid behavior, and the crossover temperature where Kondo screening sets in, respectively.  
Blue and red broken lines represent a crossover from the 4$f$-delocalized FSs to the 4$f$-localized FSs in the SDW-QCP model and the LQC model, respectively.  
}
\end{figure}

The purpose of this paper is an investigation of the FS variation around the QCP of CeRu$_2$(Si$_{1-x}$Ge$_{x}$)$_2$ by soft X-ray angle-resolved photoelectron spectroscopy (ARPES).  
To avoid a magnetic reconstruction of FSs, the ARPES measurements were performed in the temperature range $T_\mathrm{N}<T \leq T_{0}$.  
As a consequence, we have checked which of two scenarios for the quantum criticality of the HF metals is appropriate; the spin-density-wave (SDW) QCP model or the local quantum criticality (LQC) model.  
As illustrated in a schematic diagram in Fig.~\ref{PD},  
a FS crossover is expected just around the QCP in the LQC model, while it is unnecessary in the SDW-QCP model \cite{GEGENWART,SI1,SI2}.   
It is still controversial which scenario is appropriate in the HF metals \cite{KUCHLER1,PASCHEN,SCHRODER,KUCHLER2,SILHANEK,KUCHLER3,KADOWAKI}, 
and the direct observation of the FS reconstruction at the QCP has never been successful.  
ARPES experiments were previously performed to investigate the FSs of CeRu$_2$Si$_2$ and its related compounds in the VUV (20-200 eV) \cite{DENLINGER} and the soft X-ray (700-860 eV) regions \cite{YANO}.  
While Denlinger \textit {et al.} \cite{DENLINGER} could not  observe a clear difference of FSs between CeRu$_2$Si$_2$ and LaRu$_2$Si$_2$, Yano \textit {et al.} \cite{YANO} claimed the observation of 4$f$-delocalized FSs for CeRu$_2$Si$_2$ and 4$f$-localized FSs for CeRu$_2$Ge$_2$ even in the paramagnetic state.  
Therefore, it is expected that soft X-ray ARPES experiments for CeRu$_2$(Si$_{1-x}$Ge$_{x}$)$_2$ can verify the existence of the FS crossover around the QCP of this system.  
In order to enhance the photoemission signals of the Ce 4$f$ electrons, we have performed ARPES experiment in the Ce $3d$$\rightarrow$4$f$ resonance energy region and successfully observed the strongly $c$-$f$ hybridized FS both for CeRu$_2$Si$_2$ and CeRu$_2$(Si$_{0.82}$Ge$_{0.18}$)$_2$, below and above $x_{c} = 0.07$, respectively.  
ARPES was also measured for LaRu$_2$Si$_2$ as a reference material where the 4$f$ electrons do not participate in the FS formation.  


All the measured samples were single crystals grown in the procedures described in Ref.~\onlinecite{SUGI}.  
ARPES measurements were performed at beamline BL23SU of SPring-8.  
The energy resolution was 120 meV for CeRu$_2$Si$_2$ and 160 meV for CeRu$_2$(Si$_{0.82}$Ge$_{0.18}$)$_2$ at $h\nu$ = 881 eV, and 170 meV at  $h\nu$ = 770-870 eV.    
Clean sample surfaces were obtained by {\it in situ} cleaving along [001] surface.  
The sample temperature was 20 K, comparable to $T_{0}$ of CeRu$_2$Si$_2$ and above $T_{\mathrm{N}} \sim$ 8 K of CeRu$_2$(Si$_{0.82}$Ge$_{0.18}$)$_2$.  

 
\begin{figure}
\includegraphics[width=7.8cm]{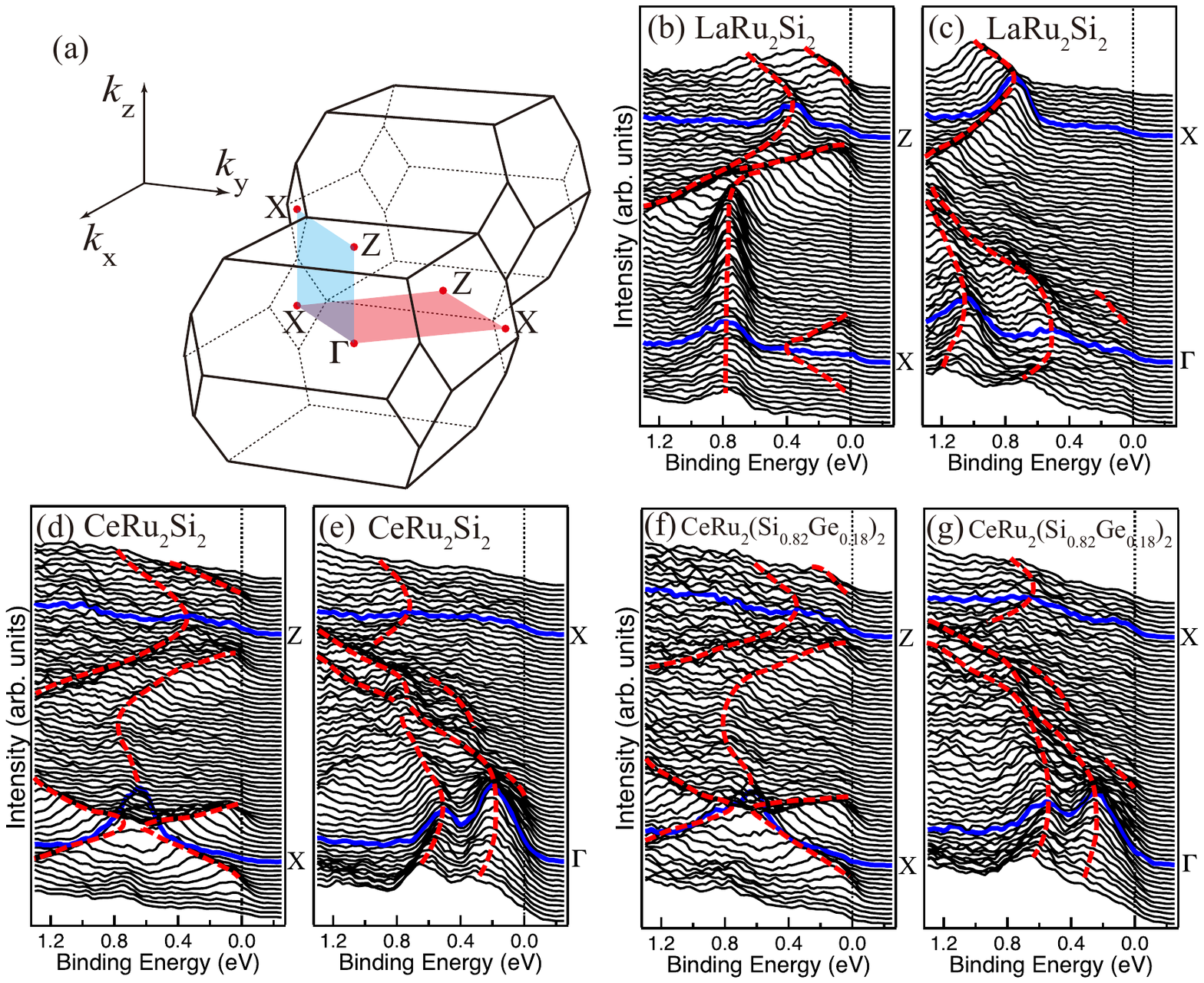}%
\caption{\label{EDC}
(color online)
(a) Brillouine zone (BZ) for CeRu$_2$(Si$_{1-x}$Ge$_{x}$)$_2$.  
Red and blue planes signify the measured plane by the angle-scanning and the $h\nu$-scanning measurements, respectively.  
Off-resonance ARPES spectra of LaRu$_2$Si$_2$ along the X-Z (b) and $\Gamma$-X (c) directions, 
those of CeRu$_2$Si$_2$ along the X-Z (d) and $\Gamma$-X (e) directions, 
and those of CeRu$_2$(Si$_{0.82}$Ge$_{0.18}$)$_2$ along the X-Z (f) and $\Gamma$-X (g) directions.  
Red broken lines indicate dispersive energy bands.   
}
\end{figure}

Figures~\ref{EDC}(b) and \ref{EDC}(c), \ref{EDC}(d) and \ref{EDC}(e), \ref{EDC}(f) and \ref{EDC}(g) show ARPES spectra of the valence bands of LaRu$_2$Si$_2$, CeRu$_2$Si$_2$, and CeRu$_2$(Si$_{0.82}$Ge$_{0.18}$)$_2$, respectively, along the X-Z and $\Gamma$-X directions of the Brillouine zone (BZ) illustrated in Fig.~\ref{EDC}(a), obtained from the angle-scanning measurements.  
The angle-scanning measurements were performed at photon energies $h\nu$ corresponding to the $\Gamma$-X-Z plane, i.e., $h\nu$ = 765, 860, and 855 eV for LaRu$_2$Si$_2$, CeRu$_2$Si$_2$, and CeRu$_2$(Si$_{0.82}$Ge$_{0.18}$)$_2$, respectively, 
which were chosen by assuming the inner potential value of 12 eV \cite{DENLINGER,YANO}.  
These cases are called {\it off-resonance} hereafter.  
As indicated by the red broken lines, the observed dispersions of energy bands are quite similar between CeRu$_2$Si$_2$ and CeRu$_2$(Si$_{0.82}$Ge$_{0.18}$)$_2$.   
On the other hand, the energy dispersions of LaRu$_2$Si$_2$ are remarkably different from those of CeRu$_2$Si$_2$: For example, the energy positions of the band structures at the $\Gamma$ point are located on the higher binding energy side in LaRu$_2$Si$_2$ than in CeRu$_2$Si$_2$.
The difference originates from the participation of the Ce 4$f$ electrons in the energy band formation, and thus it suggests the delocalized nature of the 4$f$ electrons both in CeRu$_2$Si$_2$ and CeRu$_2$(Si$_{0.82}$Ge$_{0.18}$)$_2$.  

\begin{figure}
\includegraphics[width=8.0cm]{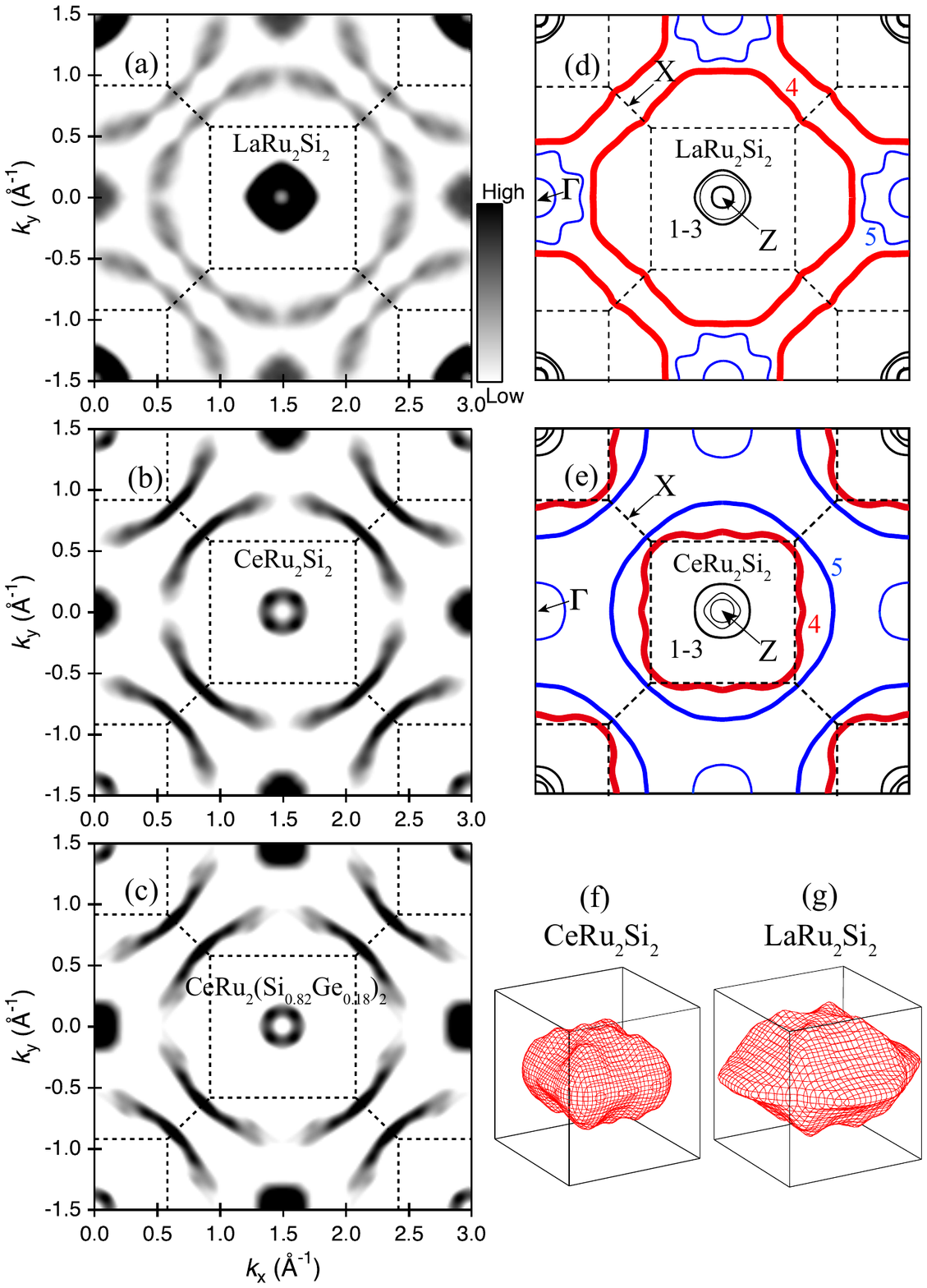}%
\caption{\label{angFS}
(color online)
Off-resonance FS images of LaRu$_2$Si$_2$ (a), CeRu$_2$Si$_2$ (b), and CeRu$_2$(Si$_{0.82}$Ge$_{0.18}$)$_2$ (c) in the $k_\mathrm{x}$-$k_\mathrm{y}$ plane, 
compared with calculated FSs of LaRu$_2$Si$_2$ (d) and CeRu$_2$Si$_2$ (e) \cite{YAMAGAMI2}.  
In (d) and (e), numbers designate the label of each FS, the same as used in Refs.~\onlinecite{DENLINGER} and \onlinecite{YANO}, and red and blue lines represent the hole FS of band 4 and the electron FS of band 5, respectively.  
BZ is illustrated by broken lines.  
Calculated 3D images of the hole FS of band 4 of CeRu$_2$Si$_2$ (f) and LaRu$_2$Si$_2$ (g) around the $Z$ point (body center) \cite{YAMAGAMI2}.  
}
\end{figure}

Figures~\ref{angFS}(a)-\ref{angFS}(c) display off-resonance FS images of LaRu$_2$Si$_2$, CeRu$_2$Si$_2$, and CeRu$_2$(Si$_{0.82}$Ge$_{0.18}$)$_2$, respectively, 
represented by the intensities of the ARPES spectra integrated near Fermi energy ($E_\mathrm{F}$) as a function of momenta ($k_x$,$k_y$). 
For comparison, the calculated FSs of LaRu$_2$Si$_2$ and CeRu$_2$Si$_2$ in the $k_x$-$k_y$ plane \cite{YAMAGAMI2} are shown in Figs.~\ref{angFS}(d) and \ref{angFS}(e), respectively.    
The numbers 1-4 and 5 designate the hole FSs and the electron FS, respectively.  
Figures~\ref{angFS}(f) and \ref{angFS}(g) show the calculated three-dimensional (3D) images of the large hole FS of band 4 of CeRu$_2$Si$_2$ and LaRu$_2$Si$_2$, respectively, indicating a significant volume variation due to participation of the Ce 4$f$ electrons in the FS formation \cite{YAMAGAMI2}.   
A large FS surrounding the Z point was experimentally detected in every compound.  
While the experimental large FS of LaRu$_2$Si$_2$ has a square shape and agrees well with the calculated hole FS of band 4, 
the experimental large FS of CeRu$_2$Si$_2$ has a circular shape and agrees well with the calculated electron FS of band 5.  
The large FS of CeRu$_2$(Si$_{0.82}$Ge$_{0.18}$)$_2$ is almost identical to that of CeRu$_2$Si$_2$.  
However, the above assignment for the Ce compounds is not conclusive because the hole FS of band 4 shown in Fig.~\ref{angFS}(e) is not observed both in CeRu$_2$Si$_2$ and CeRu$_2$(Si$_{0.82}$Ge$_{0.18}$)$_2$.    
Because the hole FS of band 4 is considered to be the heaviest FS branch due to the largest contribution of the Ce 4$f$ electrons \cite{YAMAGAMI2,AOKI}, the absence of this FS branch may deny the delocalized nature of the Ce 4$f$ electrons.   

\begin{figure}
\includegraphics[width=8.0cm]{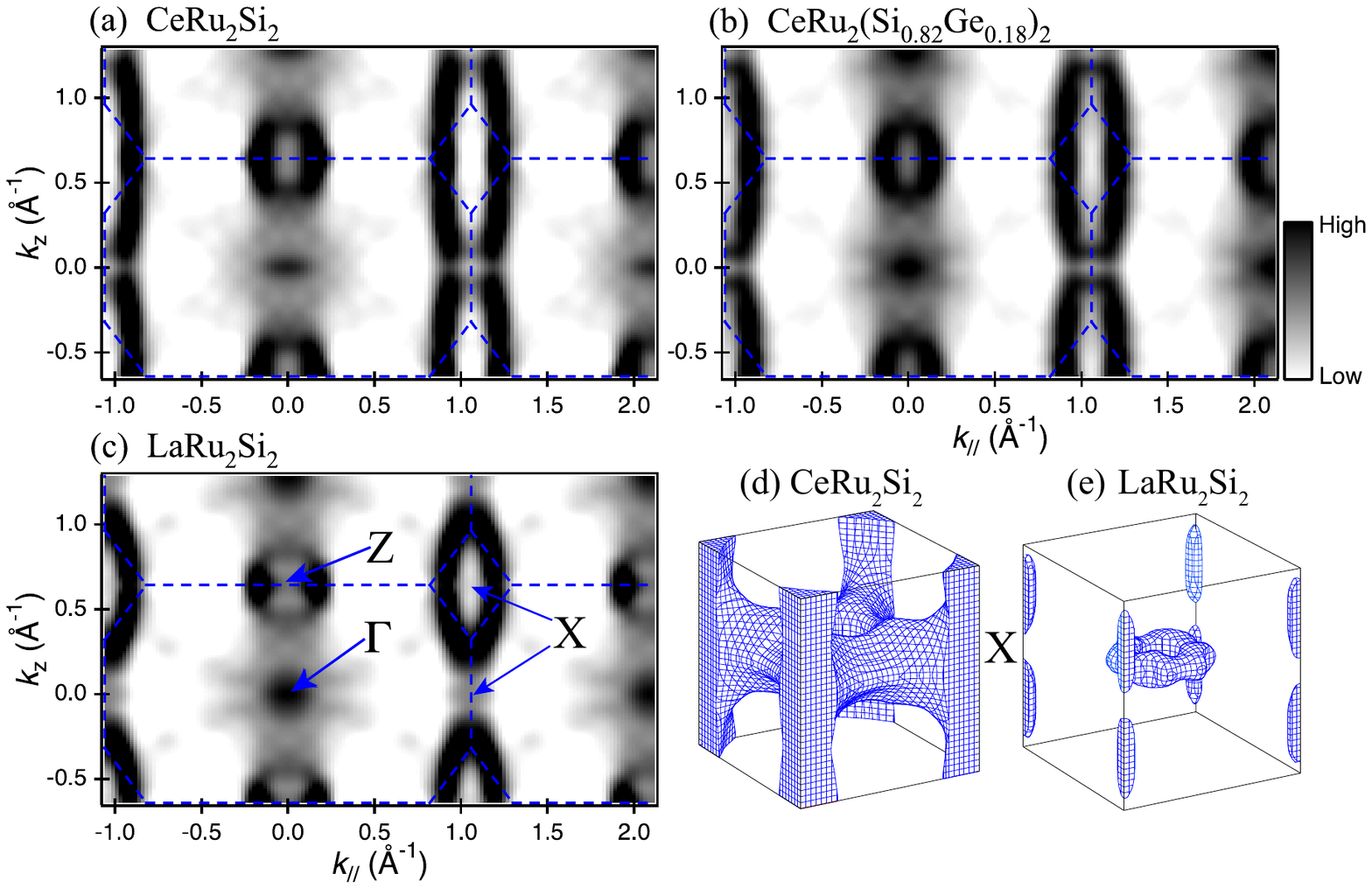}%
\caption{\label{kzFS}
(color online)
FS images of CeRu$_2$Si$_2$ (a), CeRu$_2$(Si$_{0.82}$Ge$_{0.18}$)$_2$ (b), and LaRu$_2$Si$_2$ (c) in the $k_\mathrm{xy}$-$k_\mathrm{z}$ plane.    
BZ is indicated by blue broken lines.  
Calculated 3D images of the electron FS of band 5 of CeRu$_2$Si$_2$ (d) and LaRu$_2$Si$_2$ (e) around the $\Gamma$ point (body center) \cite{YAMAGAMI2}.  
}
\end{figure}

Figures~\ref{kzFS}(a)-\ref{kzFS}(c) show FS images in the $k_\mathrm{xy}$-$k_\mathrm{z}$ plane of CeRu$_2$Si$_2$, CeRu$_2$(Si$_{0.82}$Ge$_{0.18}$)$_2$, and LaRu$_2$Si$_2$, respectively, obtained from the $h\nu$-scanning ARPES measurements (750-870 eV).  
The calculated 3D images of the electron FS of band 5 of CeRu$_2$Si$_2$ and LaRu$_2$Si$_2$ are shown in Figs.~\ref{kzFS}(d) and \ref{kzFS}(e), respectively \cite{YAMAGAMI2}.   
A noticeable difference between CeRu$_2$Si$_2$ and LaRu$_2$Si$_2$ is seen in the FS along the vertical X-X line;  
it has a connected tube-like shape in CeRu$_2$Si$_2$, while it has a closed ellipsoidal shape in LaRu$_2$Si$_2$.  
The observed difference coincides with the difference in the calculated FS between Figs.~\ref{kzFS}(d) and \ref{kzFS}(e)  \cite{YAMAGAMI2}.    
Therefore, this difference gives a good criterion whether the Ce 4$f$ electrons participate in the FS formation, 
and the present result have confirmed that the FSs of CeRu$_2$Si$_2$ are obviously different from the FSs of LaRu$_2$Si$_2$.  
The FS of CeRu$_2$(Si$_{0.82}$Ge$_{0.18}$)$_2$ along the X-X line has a tube-like shape as in CeRu$_2$Si$_2$, indicating that the 4$f$ electrons contribute to the energy-band formation in CeRu$_2$(Si$_{0.82}$Ge$_{0.18}$)$_2$.     
A remaining problem is that the difference between CeRu$_2$Si$_2$ and LaRu$_2$Si$_2$, which the large FS surrounding Z point should be composed of bands 4 and 5 in CeRu$_2$Si$_2$ and only of band 4 in LaRu$_2$Si$_2$, is not clearly observed. 

\begin{figure}
\includegraphics[width=8.0cm]{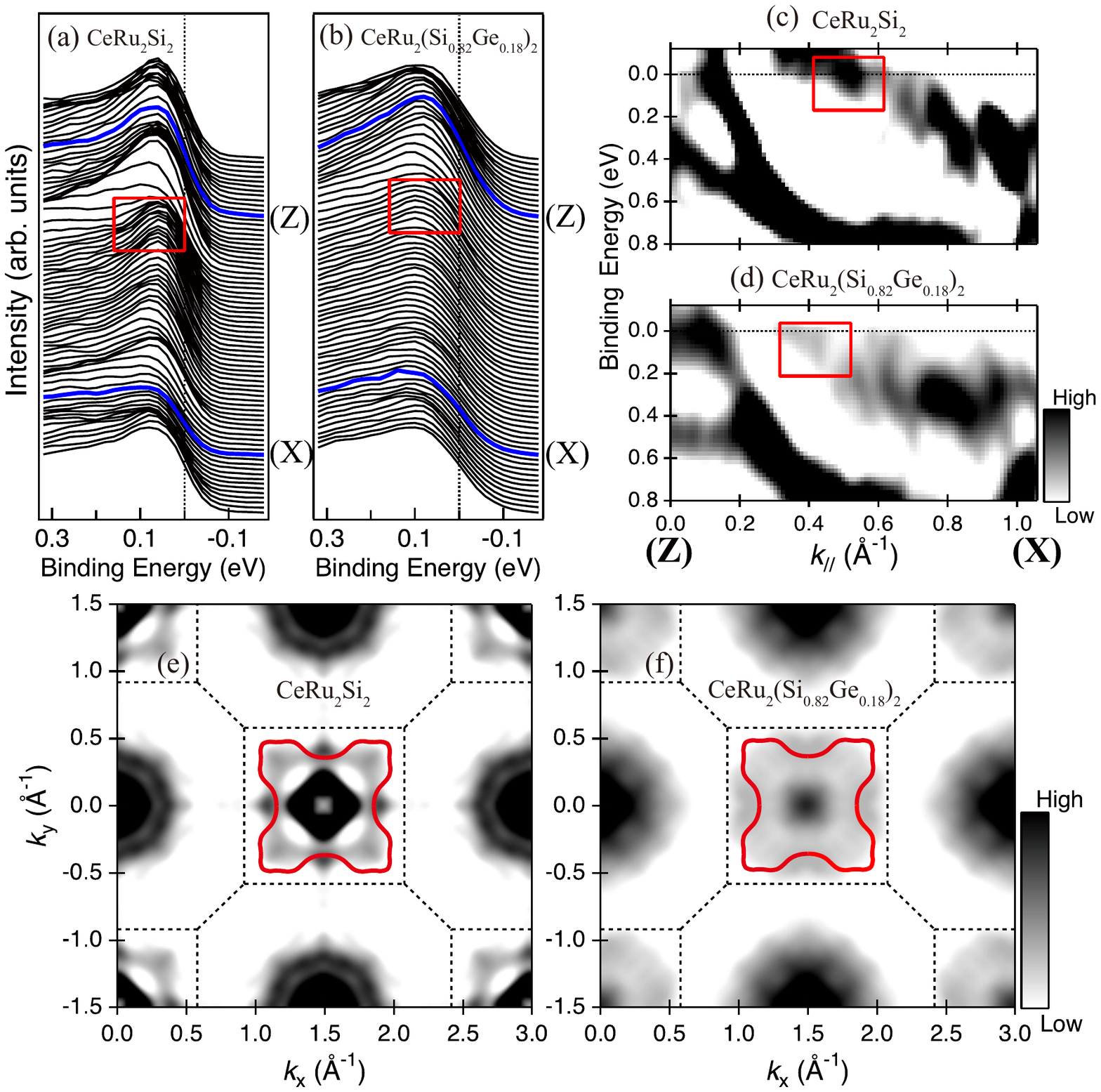}%
\caption{\label{ONRES}
(color online)
On-resonance ARPES spectra of CeRu$_2$Si$_2$ (a) and CeRu$_2$(Si$_{0.82}$Ge$_{0.18}$)$_2$ (b) along the line corresponding to the X-Z direction.  
Images of the band dispersion near $E_\mathrm{F}$ obtained from the on-resonance ARPES spectra of CeRu$_2$Si$_2$ (c) and CeRu$_2$(Si$_{0.82}$Ge$_{0.18}$)$_2$ (d).    
Red squares in (a)-(d) signify the $E_\mathrm{F}$-crossing point of the strongly $c$-$f$ hybridized  bands.  
On-resonance FS images in the $k_\mathrm{x}$-$k_\mathrm{y}$ plane of CeRu$_2$Si$_2$ (e) and CeRu$_2$(Si$_{0.82}$Ge$_{0.18}$)$_2$ (f),   
compared with the calculated hole FS of band 4 at $k_\mathrm{z} = 0.3 \pi/c$ away from the $\Gamma$-Z-X plane, represented by red lines \cite{YAMAGAMI2}.  
}
\end{figure}

The reason why the strongly $c$-$f$ hybridized hole FS is not observed in the off-resonance ARPES might be that the observed FS originate from the bands of major contribution from Ru 4$d$ electrons.  
Therefore, angle-scanning measurements were performed at the Ce $3d$$\rightarrow$4$f$ resonant energy of $h\nu$ = 881eV for CeRu$_2$Si$_2$ and CeRu$_2$(Si$_{0.82}$Ge$_{0.18}$)$_2$ to enhance the signals of the 4$f$ electrons.  
This case is called {\it on-resonance} hereafter.  
The resonance energy corresponds to  $k_\mathrm{z} = 0.29 \pi/c$ and $0.37 \pi/c$ for CeRu$_2$Si$_2$ and CeRu$_2$(Si$_{0.82}$Ge$_{0.18}$)$_2$, respectively.  
Figures~\ref{ONRES}(a) and \ref{ONRES}(b) show on-resonance ARPES spectra of CeRu$_2$Si$_2$ and CeRu$_2$(Si$_{0.82}$Ge$_{0.18}$)$_2$, respectively.  
The spectral feature around 0.1eV, which is remarkably enhanced in the on-resonance spectra, is peaky near the Z point but indistinct near the X point, indicating that the resonance enhancement is weak near the X point.  
In addition to the $E_\mathrm{F}$-crossing bands which are the same as in the off-resonance spectra in Figs.~\ref{EDC}(d) and \ref{EDC}(f), another $E_\mathrm{F}$-crossing is observed as indicated by the red squares in Figs.~\ref{ONRES}(a) and \ref{ONRES}(b).  
In order to make the dispersive bands near $E_\mathrm{F}$ clear, we have mapped images of secondary derivatives of the ARPES spectra where integrated intensities are normalized in momentum distribution curves, as shown in Figs.~\ref{ONRES}(c) and \ref{ONRES}(d). 
One can see that additional bands, which are hardly discernible in the off-resonance spectra, cross $E_\mathrm{F}$ around the midpoint of the (X)-(Z) line, designated by red squares.  
Since the additional bands appear only in the on-resonance images, they are strongly $c$-$f$ hybridized  bands.  
Figures~\ref{ONRES}(e) and \ref{ONRES}(f) display on-resonance FS images of CeRu$_2$Si$_2$ and CeRu$_2$(Si$_{0.82}$Ge$_{0.18}$)$_2$, respectively. 
Surprisingly, the large circular FS surrounding the Z point in the off-resonance images becomes invisible in the on-resonance images.  
This might be due to the weakness of the resonance enhancement near the X point mentioned above. 
In addition, another FS appears just inside the square BZ boundary surrounding the Z point both in CeRu$_2$Si$_2$ and CeRu$_2$(Si$_{0.82}$Ge$_{0.18}$)$_2$.  
This FS corresponds to the $E_\mathrm{F}$-crossing indicated by the red squares in Figs.~\ref{ONRES}(c) and \ref{ONRES}(d).  
The size and shape of this FS are comparable to those of the calculated hole FS of band 4 at $k_\mathrm{z} = 0.3 \pi/c$ away from the $\Gamma$-Z-X plane, shown by red lines in Figs.~\ref{ONRES}(e) and \ref{ONRES}(f).   
Therefore, the FS observed only in the on-resonance images is attributed to the strongly $c$-$f$ hybridized heavy-quasi-particle FS.  

The phase diagram in Fig.~\ref{PD} indicates that, in the LQC model, the FS reconstruction is expected to occur when one crosses the critical composition $x_c$ at a temperature well below $T_{0}$, even above  $T_\mathrm{N}$.   
The present result demonstrates that a discontinuous change of FS does not exist near the critical composition and the 4$f$-delocalized FS regime is extended beyond the critical composition, although the measurements were performed at a temperature comparable to $T_{0}$.  
When the temperature is lowered from $\sim T_{0}$ down to the temperature just above $T_{N}$, it is natural to think that the Ce 4$f$ electrons keep their itinerancy without a phase transition. 
Therefore, the absence of the FS change near the critical composition is strongly implied above  $T_{N}$, and it can be claimed that the SDW-QCP model is suitable for CeRu$_2$(Si$_{1-x}$Ge$_{x}$)$_2$ rather than the LQC model.

In conclusion, we have shown that the Ce 4$f$ electrons in the paramagnetic state of CeRu$_2$(Si$_{1-x}$Ge$_{x}$)$_2$ are itinerant and participate in the FS formation at both side of critical composition $x_c$ = 0.07.   
The absence of the clear change of the FSs across the critical composition indicates that the SDW-QCP model is more likely than the LQC model in CeRu$_2$(Si$_{1-x}$Ge$_{x}$)$_2$.  
In future studies, the crossover below the magnetic-ordering temperature should be examined both at the paramagnetic-antiferromagnetic and antiferromagnetic-ferromagnetic boundaries of CeRu$_2$(Si$_{1-x}$Ge$_{x}$)$_2$ by Ce $3d$$\rightarrow$4$f$ resonant ARPES.

\begin{acknowledgments}
We thank J. Otsuki for useful discussions.   
This work was performed under the Proposal No. 2008A3822 at SPring-8, supported by the Grant-in-Aid for Scientific Research on Innovative Areas of the MEXT, Japan. 
\end{acknowledgments}




\begin{references}




\bibitem{GEGENWART}
  P. Gegenwart {\it et al.},
  Nature Phys. {\bf 4}, 186 (2008).


 

\bibitem{BESNUS}
  M.J. Besnus {\it et al.},
  Solid State Commun. {\bf 55}, 779 (1985).





\bibitem{LOIDL}
  A. Loidl {\it et al.},
  Physica B. {\bf 156-157}, 794 (1989).


\bibitem{FISHER}
  R.A. Fisher {\it et al.},
  J. Low Temp. Phys. {\bf 84}, 49 (1991).


\bibitem{HAEN}
  P. Haen {\it et al.},
  Physica B. {\bf 259-261}, 85 (1999).

\bibitem{SUGI}
  M. Sugi {\it et al.},
  Phys. Rev. Lett. {\bf 101}, 056401 (2008). 
  


\bibitem{YAMAGAMI2}
  H. Yamagami {\it et al.},
  J. Phys. Soc. Jpn. {\bf 61}, 2388 (1992); J. Phys. Soc. Jpn. {\bf 62}, 592 (1993).  

\bibitem{AOKI}
  H. Aoki {\it et al.},
  Phys. Rev. Lett. {\bf 71}, 2110 (1993).

 
\bibitem{KING}
  C.A. King {\it et al.},
  Physica B. {\bf 171}, 161 (1991).

\bibitem{IKEZAWA}
  H. Ikezawa {\it et al.},
  Physica B. {\bf 237-238}, 210 (1997).

\bibitem{YAMAGAMI1}
  H. Yamagami {\it et al.},
  J. Phys. Soc. Jpn. {\bf 63}, 2290 (1992).



\bibitem{SI1}
  Q. Si {\it et al.},
  Nature {\bf 413}, 804 (2001).

\bibitem{SI2}
  Q. Si,
  Physica B {\bf 378-380}, 23 (2006).


\bibitem{KUCHLER1}
  R. K\"{u}chler {\it et al.},
  Phys. Rev. Lett. {\bf 91}, 066405 (2003). 

\bibitem{PASCHEN}
  S. Paschen {\it et al.},
  Nature {\bf 432}, 881 (2004).

\bibitem{SCHRODER}
  A. Schr\"{o}der {\it et al.},
  Nature {\bf 407}, 351 (2000).

\bibitem{KUCHLER2}
  R. K\"{u}chler {\it et al.},
  Phys. Rev. Lett. {\bf 93}, 096402 (2004). 
  
\bibitem{SILHANEK}
  A.V. Silhanek {\it et al.},
  Phys. Rev. Lett. {\bf 96}, 206401 (2006). 

\bibitem{KUCHLER3}
  R. K\"{u}chler {\it et al.},
  Phys. Rev. Lett. {\bf 96}, 256403 (2006). 

\bibitem{KADOWAKI}
  H. Kadowaki {\it et al.},
  Phys. Rev. Lett. {\bf 96}, 016401 (2006). 







  
  
 


\bibitem{DENLINGER}
  J.D. Denlinger {\it et al.},
  J. Electron Spectrosc. Relat. Phenom. {\bf 117}, 347 (2001).

\bibitem{YANO}
  M. Yano {\it et al.},
  Phys. Rev. Lett. {\bf 98}, 036405 (2007);
  Phys. Rev.B {\bf 77}, 035118 (2008). 


 
\end{references}
\end{document}